\documentclass[aps,prl,superscriptaddress,twocolumn,showpacs]{revtex4}
\usepackage{bm}
\usepackage{amsmath}
\usepackage{amssymb}
\usepackage{textcomp}
\usepackage{graphicx}
\usepackage{color}

\usepackage{soul,xcolor}
\setstcolor{red}

\input{epsf}

\begin{document}

\title{Spin-Polaronic Effects in Electric Shuttling in a Single Molecule
Transistor with Magnetic Leads}

\author{O. A. Ilinskaya}
\email{ilinskaya@ilt.kharkov.ua} \affiliation{B. Verkin Institute
for Low Temperature Physics and Engineering of the National
Academy of Sciences of Ukraine, 47 Nauki Ave., Kharkiv 61103,
Ukraine}
\author{D. Radic}
\affiliation{Department of Physics, Faculty of Science, University
of Zagreb, Bijenicka 32, Zagreb 10000, Croatia}
\author{H. C. Park}
\email{hcpark@ibs.re.kr} \affiliation{Center for Theoretical
Physics of Complex Systems, Institute for Basic Science (IBS),
Daejeon 34051, Republic of Korea}
\author{I. V. Krive}
\affiliation{B. Verkin Institute for Low Temperature Physics and
Engineering of the National Academy of Sciences of Ukraine, 47
Nauki Ave., Kharkiv 61103, Ukraine}  \affiliation{Physical
Department, V.N. Karazin Kharkiv National University, Kharkiv
61022, Ukraine}
\author{R. I. Shekhter}
\affiliation{Department of Physics, University of Gothenburg,
SE-412 96 G{\" o}teborg, Sweden}
\author{M. Jonson}
\affiliation{Department of Physics, University of Gothenburg,
SE-412 96 G{\" o}teborg, Sweden}

\date{\today}

\begin{abstract}
Current-voltage characteristics of a spintromechanical device, in
which spin-polarized electrons tunnel between magnetic leads with
anti-parallel magnetization through a single level movable quantum
dot, are calculated. New exchange- and electromechanical
coupling-induced (spin-polaronic) effects that determine strongly
nonlinear current-voltage characteristics were found. In the
low-voltage regime of electron transport the voltage-dependent and
exchange field-induced displacement of quantum dot towards the
source electrode leads to nonmonotonic behavior of differential
conductance that demonstrates the lifting of spin-polaronic
effects by electric field. At high voltages the onset of electron shuttling results in the drop of
current and negative differential conductance, caused by
mechanically-induced increase of tunnel resistivities and exchange
field-induced suppression of spin-flips in magnetic field. The
dependence of these predicted spin effects on the oscillations
frequency of the dot and the strength of electron-electron
correlations is discussed.
\end{abstract}

\maketitle

\noindent
The ability to control the spin of electrons by electrical \cite{1,2,3}, magnetic \cite{4}
and optical \cite{5} means has generated various applications in modern physics.
Spin-controlled
nanoelectromechanics \cite{fedorets1} is a promising new direction in nanoelectronics
where the interplay of spin- and mechanical degrees of freedom can lead to
novel electron transport phenomena involving the spin and charge of electrons
that can not be observed in ``ordinary'' single electron transistors.
Electron shuttling (see Ref. \onlinecite{shuttle} and experimental papers Refs. \onlinecite{exp1,exp2,exp3})
induced by the magnetic exchange forces present in a single-electron
transistor with magnetic leads was
discussed in Ref.~\onlinecite{msh}.  Unlike in a standard
single-electron shuttle \cite{fedorets}, where the electric field
generated by a bias voltage always drives the charged quantum dot (QD)
from the source- to the drain electrode, the exchange interaction
between an electron spin in the dot and the magnetization of the source
electrode results in an attractive exchange force (in what follows
we assume that the source electrode only contains spin-up electrons and the
drain electrode only spin-down electrons). For half-metallic leads
with opposite magnetization directions (see Fig.~1) the electrical current
through the system is blocked
as long as the electron
spin-projection is conserved and one needs to induce spin flips to
lift this ``spin blockade".

\begin{figure}
\includegraphics[width=0.5\textwidth]{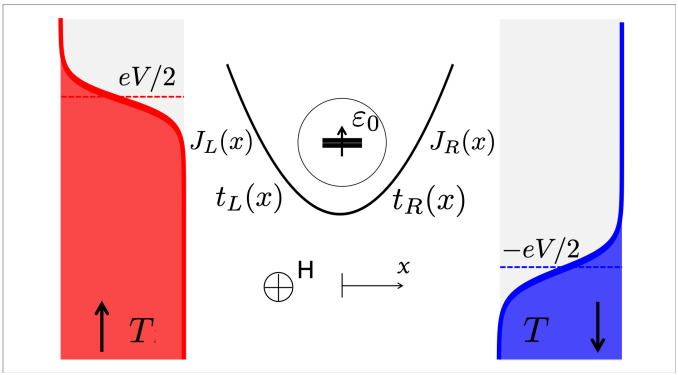}
 \caption{Sketch of the single-molecule transistor with magnetic leads discussed in the text (modelled by a single-level vibrating
  quantum dot). The dot couples to the fully and oppositely magnetized
 electrodes (spin-up electrons in the left- and spin-down electrons in the right lead) with
 position-dependent tunneling $t_{L/R}(x)$ and exchange $J_{L/R}(x)$ couplings.
 The electrons in the leads are described by equilibrium Fermi distribution functions, characterized by
 different chemical potentials $\mu_L-\mu_R=eV$ ($V$ is the bias voltage) but the same temperature $T$.
 A weak external magnetic field is applied perpendicular to
 the direction of the current and to the magnetization in the leads:
 $H\ll\Gamma/g\mu_B$ ($g$ is the gyromagnetic ratio, $\mu_B$ is the Bohr
 magneton and $\Gamma$ is level width, when the dot is located
 at the midpoint between the leads).}
\end{figure}

In a magnetic shuttle device the spin blockade
is lifted by an external magnetic field perpendicular to
both the tunneling direction and the magnetization in the leads.
This magnetic field induces oscillations between the
spin-up and spin-down projections of the spin of an electron in the dot, thus
allowing an electrical current to flow through the system. Another effect of
these oscillations is that during the time the quantum dot contains an electron in the spin-down
state it is attracted to the drain- rather than to the source electrode.
This is a necessary condition for shuttling and
magnetic shuttling is indeed realized \cite{msh} in weak magnetic
fields, $g\mu_B H \lesssim g\mu_B
H_c=\sqrt{3[\Gamma^2+(\hbar\omega)^2]}$, where $\mu_B$ is the Bohr
magneton, $g$ is the gyromagnetic ratio, $\Gamma$ is the dot-lead
tunneling coupling (the width of the QD energy level), and
$\omega$ is the angular frequency of the QD mechanical
oscillations.

The theory of magnetic shuttling developed in
Refs.~\onlinecite{msh}, \onlinecite{thsh}, and \onlinecite{ilinskaya} neglects the electric driving force. This
is a good approximation for small bias voltages when the exchange forces on the QD are
much stronger than the electric force acting on it, while for larger bias voltages both
types of forces must be taken into account. The interplay then occurring between
the electric and magnetic driving forces in a single-electron magnetic
shuttle is the subject of the present work.

In what follows we
consider a symmetric tunnel junction, i.e. the energies characterizing the tunnel couplings
of the QD (in its initial position) with the source and drain leads are
equal, $\Gamma_L=\Gamma_R\equiv \Gamma$. For simplicity we
consider a symmetric magnetic coupling as well $J_L= J_R\equiv J$
($J$ is the exchange energy in the magnetic leads). The main questions we want to answer are:
(i) is there room for spin effects in the transport regime where magnetic shuttling
is forbidden and the electric field determines the rate of electron transfer between the leads,
and (ii) if there is room, what is the signature of these effects in the current-voltage characteristic
of our spintromechanical device?

Magnetic shuttling
is possible only if there is a Coulomb blockade of tunneling \cite{ilinskaya}.
Therefore, in what follows we consider the electron transport regime
where the bias voltage is larger than the Coulomb energy, $eV> U$, so that the blockade is lifted.
(This is a different regime than considered in Ref.~\onlinecite{ilinskaya}, where it was assumed that $eV<U$).
First we calculate $I-V$ characteristics for noninteracting electrons, $U=0$, and then we
discuss how electron-electron correlations influence spin-polaronic effects.

The characteristic electric force acting on a charged QD in a
voltage biased single-electron transistor is $F_e\sim eV/d$ (here
$d$ is the distance between the source and drain electrodes). The
exchange force can be estimated as $F_m \sim J/\lambda_J$, where
$\lambda_J$ is the characteristic decay length of the exchange
interaction. These forces can act either in the same or in
opposite directions depending on the electron spin projection of
the QD electron. If it is in the spin-up state,
$\mid\uparrow\rangle$, the exchange force acts in the opposite
direction to the electric force, which pushes the QD towards the
drain electrode. If the QD electron is in the spin-down state,
$\mid\downarrow\rangle$, both forces drive the QD towards the
drain electrode. Spin-up states dominate in weak magnetic fields
($g\mu_B H\ll\Gamma$), in which spin flips are suppressed. This is
the case that will be considered in what follows. We show that
these states at low vibration frequencies, $\hbar\omega\ll\Gamma$,
lead to a nonmonotonic behavior of the differential conductance
even in the case when mechanical subsystem is stable with respect
to buildup of developed oscillations (''vibronic`` regime of
electron transport). It is evident from above considerations that
there is a critical bias voltage of the order of $eV_c\sim
(d/\lambda_J)J$ when for $V>V_c$ electric shuttling occurs. In the
shuttling regime of electron transport (periodic oscillation of
the QD with voltage-dependent amplitude) large amplitudes of dot
oscillations result in pronounced
spin-polaronic effects.

Another physical phenomenon significant for the magnetic shuttle
dynamics is the strength $U$ of electron-electron correlations in the
QD. This parameter determines the population of the doubly occupied electron
state $|2\rangle$. In the Coulomb blockade regime $U\gg T, eV$
(where $T$ is the temperature) magnetic shuttling is a possible
regime of electron transport \cite{msh} in small magnetic fields
$H\lesssim H_c$. For noninteracting electrons, $U=0$, magnetic
shuttling in symmetric junction is not realized in the whole range
of model parameters \cite{ilinskaya,LTP}. In this case, at bias voltages for which
the electric force exceeds the exchange force, electric shuttling
takes place. The transformation from the vibronic regime of electron
transport in exchange force-based spintronic transistor to
electric shuttling is another significant problem we study
in the present paper. We have shown that this transition is
manifested in a current drop and therefore in a negative differential conductance
in our spintromechanical
device. The size of the current drop increases with an increase of frequency
but saturates at a frequency of order $\hbar\omega\simeq\Gamma$.

The functionality of the magnetic shuttle device is determined by
the
interplay between three different physical processes:
(i) the
tunneling of electrons through a single-level vibrating QD, which interacts with
the leads by coordinate-dependent exchange and tunnel interactions,
(ii) the mechanical motion of
the movable dot, which affects the electron tunneling probabilities and
the electron energy level in the dot,
and (iii) the external magnetic field-controlled electron spin
dynamics, which influences the
mechanical motion of the dot through the exchange force,
acting on the quantum dot.

The Hamiltonian of our system consists of three different terms,
$\hat H=\hat H_l+\hat H_d+\hat H_{tun}$. Noninteracting, fully and oppositely spin-polarized
electrons in the left (L) and right (R) leads are described by the
Hamiltonian $\hat{H}_l$,
\begin{equation}\label{Hl}
\hat H_l=\sum_{k,j}\varepsilon_{k,j}a_{k,j}^\dag
a_{k,j},\qquad j=L,R=\uparrow,\downarrow.
\end{equation}
Here $a_{k,j}^\dag$ $(a_{k,j})$ is the creation (annihilation)
operator of an electron with momentum $k$ ($\varepsilon_{k,j}$ is the electron energy)
in lead $j=L, R$.

The QD Hamiltonian is 
a sum of two contributions, $\hat
H_d=\hat H_d^e+H_d^v$, which describe respectively the interacting
electron- and vibron subsystems. The vibronic subsystem is modelled
by the Hamiltonian of a harmonic oscillator,
\begin{equation}\label{Hdv}
H_d^v=\frac{p^2}{2m}+\frac{m\omega^2}{2}x^2,
\end{equation}
where the center-of-mass coordinate $x$ and the
momentum $p$ of the QD are treated as classical variables, $m$ is the mass of the QD and
$\omega$ is the angular frequency of the dot vibrations.
The electronic part reads
\begin{equation}\label{Hde}
\hat H_d^e=\sum_\sigma\varepsilon_\sigma(x) c_\sigma^\dag
c_\sigma -\Omega_H\left(c_\uparrow^\dag
c_\downarrow+c_\downarrow^\dag c_\uparrow\right)+Uc_\uparrow^\dag
c_\uparrow c_\downarrow^\dag c_\downarrow,
\end{equation}
where $c^{\dagger}_{\sigma}$ ($c_{\sigma}$) is the creation
(annihilation) operator of an electron with spin projection
$\sigma$ in the QD, $\varepsilon_{\sigma}(x)=\varepsilon_0-e E
x-(\sigma/2)J(x)$ [with $\sigma = \uparrow,\downarrow =+,-$] is
the spin- and position-dependent energy of the Zeeman-split dot
level $\varepsilon_0$ in the exchange field $J(x)=J_L(x)-J_R(x)$.
Here $J_{L/R}(x)=J_{L/R}\exp(\mp x/\lambda_J)$ is the interaction
energy due to magnetic exchange interactions between
the magnetization in the ferromagnetic leads, $L/R$, and a unit
spin on the dot, $\lambda_J$ is the characteristic decay length of
the exchange interaction, $E$ is the strength of the electric field between the leads, and $e$ is the
electron charge. In Eq.~(\ref{Hde}) $\Omega_H/\hbar=g\mu_B
H/(2\hbar)$ is the Larmor frequency of electron precession in the
external magnetic field $H$, directed perpendicular to the
antiparallel magnetizations in the leads, $g$ is the gyromagnetic
ratio, $\mu_B$ is the Bohr magneton; $U$ is the Coulomb repulsion
energy.

In the  standard tunneling Hamiltonian,
\begin{equation}\label{Ht}
\hat H_{tun}=t_L(x)\sum_k c^\dag_\uparrow a_{k,L}+t_R(x)\sum_k
c^\dag_\downarrow a_{k,R}+\text{h.c.} ,
\end{equation}
the coordinate dependence of the tunneling amplitudes,
$t_{L,R}(x)$ is taken into account. This $x$-dependence is
modelled by the one-parameter exponential function
$t_{L/R}(x)=t_{L/R}\exp(\mp x/2\lambda)$, where $\lambda$ is the
characteristic tunneling length, and the dot coordinate $x$ is
measured from the isolated dot position in the center of the gap
between the electrodes.

The classical nanomechanics of the shuttle vibrations
(the time-dependent displacement $x(t)$ of the dot center-of-mass
coordinate) can be described by Newton's equation for the
oscillator with a spin- and displacement-dependent exchange force,
which is strongly nonlinear in $x(t)$ and nonlocal in
time \cite{ilinskaya}. Here, we derived this equation by taking into
account the bias voltage-dependent electric force acting on the dot
\begin{eqnarray}\label{x}
\ddot x+\omega^2
x=-\frac{1}{2m}\frac{\partial J(x)}{\partial x}\left[\rho_\uparrow\{x(t')\} -
\rho_\downarrow\{x(t')\}\right]\\ \nonumber
+\frac{eV}{md}[\rho_2\{x(t')\}-
\rho_0\{x(t')\}]+\frac{eV}{md}.
\end{eqnarray}

The probabilities $\rho_j$ ($j=0, \uparrow, \downarrow, 2$;
$\sum\rho_j=1$) are the diagonal matrix elements of the QD density
operator $\hat{\rho}_d$ in the four-dimensional Fock space of a single-level QD:
$\rho_0=\langle 0|\hat{\rho}_d|0\rangle$,
$\rho_{\uparrow}=\langle\uparrow|\hat{\rho}_d|\uparrow\rangle$, $\rho_{\downarrow}=
\langle \downarrow|\hat{\rho}_d|\downarrow\rangle$,
$\rho_2=\langle 2|\hat{\rho}_d|2\rangle$.
Here $|0\rangle$  is the ground state (empty QD), the singly
occupied electron states are $|\uparrow(\downarrow)\rangle=
c^{\dagger}_{\uparrow(\downarrow)}|0\rangle$, and the doubly
occupied electron state is
$|2\rangle=c^{\dagger}_{\uparrow}c^{\dagger}_{\downarrow}|0\rangle$.

The probabilities $\rho_j$ and the non-diagonal matrix element
$\rho_{\uparrow\downarrow}= \langle
\uparrow|\hat{\rho}_d|\downarrow\rangle =
\rho_{\downarrow\uparrow}^{\ast}$ obey the set of linear
differential equations derived in
Refs.~\onlinecite{ilinskaya,LTP}. The main approximation in the
derivation of these equations (see Eqs. (A.1)--(A.9) in
Appendix A of Ref.~\onlinecite{LTP}) is that perturbation theory is applied, using
the level width as a small parameter, $\Gamma/ \max\{T, eV\}\ll 1$. This allows one to (i)
represent the total density operator of our system as a product
of the desired QD density operator $\hat{\rho}_d$ and the equilibrium
density matrices of the leads \cite{novotny}, and to (ii) trace out the electron degrees of
freedom in the leads and make the Liouville--von Neumann
equation for the reduced density operator local in time. In our case,
when the electric force is taken into account, the set of equations
derived in the Appendix of Ref.~\onlinecite{LTP} is slightly
changed, viz. $\varepsilon_0$ is replaced by $\varepsilon_0-e Ex$.
In what follows we will analyze the resulting coupled nonlinear (and nonlocal in time) mechanical
equation and the set of linear kinetic equations numerically.

Now we derive an analytic expression for the average electric current,
$I=Tr(\hat{\rho}\hat{I})$, where the full density operator
$\hat{\rho}$ for sequential electron tunneling is factorized as
$\hat{\rho}\simeq\hat{\rho}_d \hat{\rho}_l$ (here
$\hat{\rho}_l=\hat{\rho}_{L}\hat{\rho}_{R}$ is the equilibrium
density matrix of the source and drain electrodes). The electric tunnel current
operator in the source/drain lead ($j=L/R$) takes
the standard form
\begin{equation} \label{Jj}
 \hat{I}_j=-ie\sum_k t_j(x)(c_j^{\dagger}a_{k,j}-a_{k,j}^{\dagger}c_j)  .
\end{equation}
We calculate the average current in the adiabatic transport regime,
where the dot coordinate $x(t)$ can be considered to be a slowly varying
function of time. In this case we can neglect the dynamics of $x(t)$
and treat the dot coordinate as a parameter when deriving $I_j\{x(t)\}$.
The current-voltage dependencies in the asymptotic
regime of QD oscillations are obtained by time-averaging of the
current during one period $\Delta t=2\pi/\omega$,
\begin{equation}\label{JV}
I_j(V)= \frac{1}{\Delta t}\int_{t_s}^{t_s+\Delta t}\; I_j\{x(t)\} dt.
\end{equation}
Here the time $t_s$ is chosen large enough for the QD oscillation amplitude to have reached a stationary value.
With this constraint the averaging does not depend on the specific choice of $t_s$.

In the adiabatic limit the average current can be expressed in terms of
matrix elements of the QD density operator. To obtain this expression,
we calculate the trace in the equation for the electric current in the interaction representation. By
using the integral form of the Liouville--von Neumann equation, one easily
finds the result
\begin{equation}\label{J1}
I_j=-i\,Tr\left\{\widetilde{I}_j(t)\int_{-\infty}^t dt^\prime\left[\widetilde{H}_{tun}(t^\prime),\widetilde{\rho}(t^\prime)\right]\right\}\,,
\end{equation}
where all operators are in the interaction representation, $\widetilde{I}(t)=\exp{(i\hat H_0 t)}\hat I \exp{(-i\hat H_0 t)}$, $\hat H_0=\hat H_l+\hat H_d$.
Now Eq.~(\ref{J1}) can be rewritten in a more convenient form for
further calculations
\begin{equation}\label{J2}
\begin{split}
I_j=-i\int_0^\infty d\tau\,Tr&\left\{\left(e^{i\hat H_0 \tau}\hat {I}_j
e^{-i\hat H_0 \tau}\hat H_{tun}\right.\right.\\
&\left.\left.-\hat H_{tun}e^{i\hat H_0 \tau}\hat I_j e^{-i\hat H_0
\tau}\right)\hat\rho\right\}.\\
\end{split}
\end{equation}
We calculate the current in the source lead ($I_L=-I_R\equiv I$). By inserting Eq.~(\ref{Ht}) for
the tunneling Hamiltonian and Eq.~(\ref{Jj}) for the electrical current operator into Eq.~(\ref{J2})
and noting that $\exp{(i\hat H_l \tau)}\, a_{k,L} \exp{(-i\hat
H_l \tau)}=\exp{(-i\epsilon_k\tau)}a_{k,L}$, one finds that
\begin{widetext}
\begin{equation}\label{J3}
\begin{split}
I=-e\int_0^\infty d\tau\sum_k\,t_L^2(x)\,e^{i\epsilon_k\tau}\,Tr&\left\{\left(-a_{k,L}^{\dag}e^{i\hat H_d \tau} c_\uparrow e^{-i\hat H_d \tau}c_\uparrow^{\dag}a_{k,L} \right.\right.\\
&\left.\left.+c_\uparrow^\dag a_{k,L} a_{k,L}^\dag e^{i\hat H_d \tau} c_\uparrow e^{-i\hat H_d \tau}\right) \hat\rho_l \hat\rho_d\right\}+\text{h.c.}\\
\end{split}
\end{equation}
\end{widetext}

In our approximation the traces over the electronic degrees of
freedom in the leads and in the dot can be evaluated
separately. For the electrons in the leads one finds that
$Tr\left(\hat\rho_l a_{k,L}^\dag a_{k,L}\right)=f_L(\epsilon_k)$
and $Tr\left(\hat\rho_l a_{k,L}
a_{k,L}^\dag\right)=1-f_L(\epsilon_k)$, where $f_L(\varepsilon)$
is the Fermi distribution function for chemical potential $\mu_L$
and temperature $T$. It follows that the equation for the current
takes the form\begin{widetext}
\begin{equation}
\begin{split}\label{J4}
I=-e\int_0^\infty d\tau\sum_k\,t_L^2(x)\,e^{i\epsilon_k\tau}\,&\left\{-f_L(\epsilon_k)\,Tr\left(\hat\rho_d e^{i\hat H_d \tau} c_\uparrow e^{-i\hat H_d \tau}c_\uparrow^{\dag}\right)\right.\\
&\left.+[1-f_L(\epsilon_k)]\,Tr\left(\hat\rho_d c_\uparrow^{\dag} e^{i\hat H_d \tau} c_\uparrow e^{-i\hat H_d \tau}\right)\right\}+\text{h.c.}\\
\end{split}
\end{equation}
\end{widetext}
To evaluate traces over the QD operators, it is necessary to diagonalize
the electronic part $\hat H_d^e$ of the dot Hamiltonian by the unitary transformation (see Ref.~\onlinecite{thsh})
\begin{equation}
\begin{split}
&b_1=\cos{\varphi}\; c_\uparrow - \sin{\varphi}\; c_\downarrow\,,\\
&b_2=\sin{\varphi}\; c_\uparrow + \cos{\varphi}\; c_\downarrow\,,\\
\end{split}
\end{equation}
with $\tan{2\varphi}=-2\Omega_H/J(x)$. Using the fact that the vectors
$|0\rangle$, $b_1^\dag|0\rangle$, $b_2^\dag|0\rangle$, $b_1^\dag
b_2^\dag|0\rangle$ are eigenvectors of $\hat H_d^e$ with
eigenvalues $0$, $E_+$, $E_-$, and $E_{+} + E_{-} +U$, respectively,
($E_\pm = \varepsilon_0-eEx \pm (1/2)\sqrt{4\Omega_H^2 +
J^2(x)}$), the traces in Eq.~(\ref{J4}) can easily be evaluated.
The result of the summation over $k$ reads
\begin{equation}\label{sum1}
\sum_k\,e^{i\epsilon_k\tau}f_L(\epsilon_k)=\frac{-\pi i \nu_L T e^{i\mu_L\tau}}{\sinh{[\pi T (\tau - i 0)]}}
\end{equation}
and
\begin{equation}\label{sum2}
\sum_k\,e^{i\epsilon_k\tau}[1-f_L(\epsilon_k)]=2\pi\nu_L e^{i\mu_L\tau}\delta(\tau)-\sum_k\,e^{i\epsilon_k\tau}f_L(\epsilon_k)\,,
\end{equation}
where $\delta$ denotes the Dirac delta-function and $\nu_L$ is the density of states in the source lead
which is assumed to be constant (wide-band approximation). After evaluating the integrals over $\tau$, such as
\begin{equation}\label{int}
\int_{-\infty}^{\infty}d\tau\,\frac{e^{i(\mu_L-E_+)\tau}}{\sinh{[\pi T (\tau - i 0)]}}=\frac{2i}{T}f_L(E_+)\,,
\end{equation}
we obtain the desired result for the electrical current in terms of the matrix elements of the QD density operator as
\begin{equation}\label{Ifin}
\begin{split}
I=e&\left\{-\Gamma_L(x)(\rho_\uparrow + \rho_2)+[\Gamma_L(x)f_L^+ + \Upsilon_{1L}(x)](\rho_0 + \rho_\uparrow)\right.\\
&+[\Gamma_L(x)f_L^{U,+} + \Upsilon_{1L}^U(x)](\rho_\downarrow + \rho_2)\\
&\left.+2[\Upsilon_{2L}(x)-\Upsilon_{2L}^U(x)]\text{Re}\rho_{\uparrow\downarrow}\right\}\,.\\
\end{split}
\end{equation}
Here $2f_{L,R}^{\pm}=f_{L,R}(E_{-})\pm f_{L,R}(E_{+})$ and
$2f_{L,R}^{U,\pm}=f_{L,R}(E_{-}+U)\pm f_{L,R}(E_{+}+U)$, while
\begin{eqnarray}\label{Ups-1}
\Upsilon_{1L/R}(x)&=&f_{L/R}^{-}\frac{J(x)\Gamma_{L/R}(x)}{\sqrt{J^2(x)+4\Omega_H^2}}\,,\\
\Upsilon_{1L/R}^{U}(x)&=&f_{L/R}^{U,-}\frac{J(x)\Gamma_{L/R}(x)}{\sqrt{J^2(x)+4\Omega_H^2}}\,,\\
\Upsilon_{2L/R}(x)&=&f_{L/R}^{-}\frac{\Omega_H\Gamma_{L/R}(x)}{\sqrt{J^2(x)+4\Omega_H^2}}\,,\\
\Upsilon_{2L/R}^{U}(x)&=&f_{L/R}^{U,-}\frac{\Omega_H\Gamma_{L/R}(x)}{\sqrt{J^2(x)+4\Omega_H^2}}\,.
\label{Ups-2}
\end{eqnarray}

Results of numerical calculations are presented in Figs.~2(a),
2(b), 3 and 4. In Fig.~2(a) $I-V$ curves are plotted for two
different frequencies; $\hbar\omega \simeq 0.15\,\Gamma$,
$\hbar\omega = 0.50\,\Gamma$. The currents are normalized to the
maximum (saturation) current
$I_m=(e\Gamma/2\hbar)[4\Omega_H^2/(\Gamma^2+4\Omega_H^2)]$ of a
spintronic transistor with an immobile quantum dot symmetrically
coupled to the leads (see, e.g., Ref.~\onlinecite{Zubov}). The
plots reveal a significantly non-monotonic dependence of the
electrical current on bias voltage. Figure~2(a) clearly shows that
there is a low bias-voltage region and a high bias-voltage region,
which are characterized by a qualitatively different behavior of
the current. At low biases the current grows with increasing
voltage until at some threshold voltage $V_{th}$ it almost reaches
the maximum current, $I_m$, through a symmetric junction, a small
deviation being due to the fact that occupation factors differ
slightly from 1 or 0 (cf.  Eq.(\ref{Imodel})).

\begin{figure}
\includegraphics[scale=0.5]{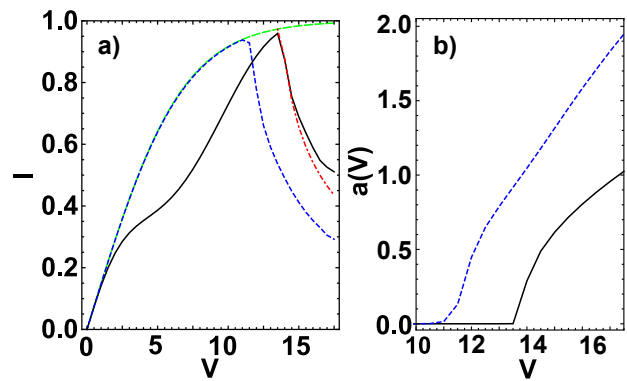}
 \caption{ a) Current through the model spintromechanical transistor shown in
Fig.~1 as a function of bias voltage for different
 dot oscillation frequencies. The current is normalized to the maximum (saturation) current of a device
 with an immobile quantum dot, $I_m=(e\Gamma/2\hbar)[4\Omega_H^2/(\Gamma^2+4\Omega_H^2)]$ (see Ref.~\onlinecite{Zubov}];
 the bias voltage is measured in units of $\Gamma$. The solid (black on-line) curve corresponds to
$\hbar\omega/\Gamma\simeq 0.15$ and the
short-dashed curve (blue on-line) to $\hbar\omega=0.5\Gamma$. Curves for higher frequencies
 (e.g., $\hbar\omega=1.5\Gamma$) practically coincide with the short-dashed curve and are not shown. The dashed-dotted curve (red on-line) is the result of a simplified semi-analytical model calculation
 (see text) and the dashed curve (green on-line) is the $I-V$ curve
 for a device with an immobile quantum dot. The plots demonstrate that in contrast to what is the case for
 an ordinary single-electron transistor, the $I-V$ characteristics of a spintromechanical
 transistor are strongly nonlinear.
 b) The dependence of amplitude $a(V)$ of the developed shuttling oscillations (normalized to the tunneling length $\lambda$)
 on bias voltage normalized by $\Gamma$. The solid curve corresponds to $\hbar\omega/\Gamma\simeq 0.15$ and the short-dashed curve (blue on-line) to $\hbar\omega/\Gamma=0.5$. It is clearly seen from the plots that low- and high-voltage parts of $I-V$ curves
 in Fig.2(a) correspond to vibronic and shuttling regimes of electron transport. Other parameters used in numerical calculations
 are: the Coulomb correlation energy $U=0$, the level energy (counted from the Fermi energy) $\varepsilon_0/\Gamma=1$, magnetic field $\Omega_H/\Gamma=0.1$,
 temperature $T/\Gamma=1.5$, exchange coupling $J/\Gamma=0.77$, the electromechanical coupling constant $\kappa=\hbar^2/(m\lambda\lambda_J\Gamma)=0.06$, where
 $\kappa$ is the coefficient multiplying r.h.s. of Eq.~(\ref{x}) written in terms of
 dimensionless variables $x$ and $t$, and the distance between the leads is set to be $d=20\lambda$.}
\end{figure}

With a further increase of bias voltage the current decreases rapidly. The magnitude of the current drop is
of the order of the maximum current. The two different regimes of electron transport are related
to two different phases of the mechanical QD oscillations. In the low-bias regime
the amplitude $a(V)$ of the QD oscillations is vanishingly small (vibronic phase). At the threshold bias
voltage the amplitude starts to grow rapidly until it reaches a value of a few $\lambda$ (see Fig.~2(b), where the
QD vibration amplitude $a(V)$ is normalized to the tunneling length $\lambda$). This is the shuttling regime
of electron transport \cite{fedorets,fedorets1}.

\begin{figure}
 \includegraphics[scale=0.7]{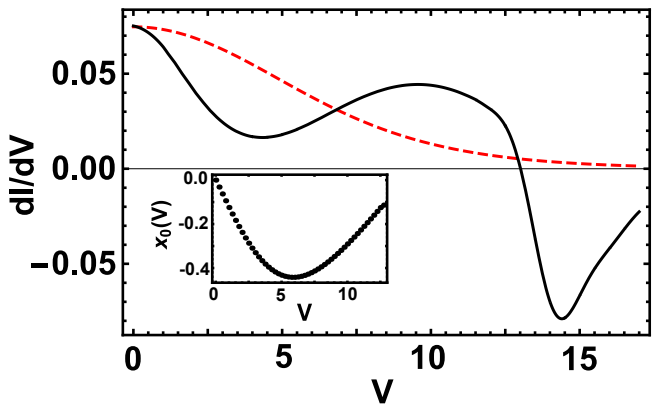}
 \caption{ Differential conductance as a function of bias voltage.
 The solid (black on-line) curve corresponds to
 $\hbar\omega/\Gamma\simeq 0.15$ and the dashed (red on-line) curve is for an immobile quantum dot.
 The oscillation of magnetic shuttle introduces the negative differential conductance in $V\gtrsim 13$.
 The inset shows the exchange and electric force-induced shift of dot coordinate $x_0(V)$. Strongly
 non-linear dependence of $x_0(V)$ on bias voltage results in non-monotonic differential conductance.
 Other parameters used in numerical calculations are the same as in Fig.~2.}
 \end{figure}

As mentioned in the Introduction, a rough estimate of the
threshold bias voltage gives $eV_{th}\sim Jd/\lambda\simeq
16\,\Gamma$ (for $J=0.77\,\Gamma$ and $ d/\lambda=20 $,
$\lambda = \lambda_J$). This is in a
reasonably good agreement ($\sim 50\%$) with numerical results for
the same parameters (see Fig.~2(a)). Notice, however, that the
threshold voltage observed numerically depends on frequency (see
Fig.~2(b)) and that this frequency dependence in the considered
spintromechanical device looks qualitatively different from the
one for shuttling of spinless electrons, in which case
$eV_{th}\propto\hbar\omega$ (see Ref.~\onlinecite{fedorets}). It
is clear from Fig.~2(b) that $V_{th}(\omega_1)\leq
V_{th}(\omega_2)$ for $\omega_2\leq\omega_1$. However, the effect
is not numerically large since $ \max(\delta V_{th}/V_{th}) <
0.2$.

Let us first explain the non-monotonic behavior of the
differential conductance, $dI/dV$, in the vibronic ''phase`` of
dot oscillations and focus on the adiabatic regime, where
$\hbar\omega\ll\Gamma$ (see Fig.~3, where the
red curve shows the voltage dependence of the differential
conductance of a non-movable quantum dot as calculated using
Eq.~(A.20)
of Ref.~\onlinecite{Zubov}). For small magnetic fields, $h\equiv
g\mu_B H \ll\Gamma$, the exchange force induces a shift, $x_0(V) <
0$, of the quantum dot in the direction of  the source electrode
if the dot is occupied by a spin-up electron. This shift depends
non-monotonically on the bias voltage (see the inset in Fig.~3)
due to the different voltage-dependencies of the electric and
magnetic forces. The electric force has a simple linear dependence
on the bias voltage while the voltage dependence of the exchange
force is more complicated. This is because it has an exponential
dependence on the voltage dependent shift $x(V)$ of the dot, $\sim
J\sinh x_0(V)$. In addition the exchange force depends on the
voltage-dependent probability for the dot to be occupied by a
spin-up electron, see Eq.~(\ref{x}).

At zero bias, $V=0$,  the probabilities
to fill the dot level with a spin-up electron from the source or a spin-down electron from the drain are equal,
so there is no net exchange force on the dot (nor is there any electric force). A finite bias voltage
favors spin-up states on the dot and therefore leads to an exchange force that acts to shift the dot
towards the source electrode, a shift that itself increases the exchange force. The exchange force is opposed by the electric force, $\propto V$,
and the elastic restoring force, $\propto \omega^2 x_0(V)$, which leads to a voltage dependent shift $x_0(V)$ for which these forces balance
each other.  The shift has a maximum value for some finite voltage and decreases if the voltage is increased further as
the electric force, which is always directed towards the drain electrode, grows in strength.
When the electric force equals the magnetic force the dot returns to its symmetric position and there is no shift (see the inset in Fig.~3).
(Note that in the vibronic phase at low frequencies this maximum shift  is always smaller than the tunneling length $\lambda$,
decreases with increasing $\omega$, and saturates at negligibly small values when $\hbar\omega\geq\Gamma$.)
As a result, the differential conductance is a non-monotonic function of voltage. This is a
special feature of a spintromechanical transistor in the adiabatic regime of dot oscillations. The effect
disappears at high oscillation frequencies $\hbar\omega\geq\Gamma$ when the voltage-induced shift of the
quantum dot is negligibly small and the $I-V$ characteristics coincide with the simple analytical dependence known for an
immobile quantum dot \cite{Zubov}.

Now we proceed to a discussion of the shuttling regime of
transport. Electric shuttling starts in the vicinity of the
threshold bias voltage for which the current almost reaches the
maximum value that can be achieved in a symmetric junction. If we
neglect small frequency-dependent variations in $V_{th}$, electric
shuttling begins exactly at the point where the electric and magnetic forces
compensate each other, the quantum dot is in its symmetrical
position and the current is maximal. At these high voltages the
dot level (in our simulations
$\varepsilon_0=1\,\Gamma$) is populated  by spin-up
electrons with almost 100\% probability. The magnetic force is
compensated and there is no threshold for electric shuttling. Is
there room for spin-induced effects in this regime of electron
transport? To answer this question we consider a simple adiabatic
model for shuttling, using the numerically evaluated dependence of
the shuttling amplitude on bias voltage, see Fig.~2(b).

In Ref.~\onlinecite{Zubov} an analytic expression for the electric current
in a spintronic transistor with an immobile quantum dot
was derived (see Eq.~(A.20) in the Appendix of the cited paper)
\begin{equation}\label{Imodel}
 I=e\frac{\Gamma_L\Gamma_R}{\Gamma_L+\Gamma_R}\frac{h^2}{\Gamma_L\Gamma_R+ h^2}(f_L^{+}-f_R^{+}).
\end{equation}
Here $h=g\mu_B H$ and $2f^{\pm}_{L,R}=f_{L,R}(\varepsilon_1)\pm f_{L,R}(\varepsilon_2)$, $\varepsilon_{1,2}=
\varepsilon_0\pm h/2$ are the Zeeman-split dot energy levels 
and $f_j(\varepsilon)$ is the Fermi distribution function
characterized by chemical potential $\mu_j$ and temperature $T$.
We generalize this simple model for our case by assuming that in
the adiabatic regime of dot oscillations one can replace the
tunneling couplings $\Gamma_j$ by the time-dependent quantities
$\Gamma_j\{x(t)\} =\exp(\mp x(t)/\lambda)$, where
$x(t)=a(V)\cos\omega t$ is the voltage-dependent amplitude of
shuttle oscillations (see Fig.~2(b)). We include also the
exchange-interaction-induced time-dependent gap in the level
splitting $\delta\varepsilon=\sqrt{h^2+4J^2\sinh^2[x(t)/\lambda]}$ in order to take into
account the effects of the exchange interaction. Now the
time-dependent current takes the form (we omit the last factor in
Eq.(\ref{Imodel}), which describes the effects of level
populations; we numerically checked that this factor is very
close to one and it does not modify the calculated $I-V$ characteristics in the shuttle regime of transport)%
\begin{equation} \label{I(t)}
 I_m(t)=\frac{e\Gamma}{2\hbar\cosh[x(t)/\lambda]}\frac{h^2}{\Gamma^2+h^2+4J^2\sinh^2[x(t)/\lambda]} .
\end{equation}
This current depends on bias voltage through the voltage-dependent
amplitude $a(V)$ of shuttling, Fig.~2(b). The current
Eq.~(\ref{I(t)}) averaged over one period of oscillation
$2\pi/\omega$ results in the desired current-voltage
characteristics $\bar{I}(V)$ plotted in Fig.~2(a) (dashed-dotted curve, red on-line)
for the case of a low frequency $\hbar\omega\simeq 0.15\,\Gamma$.

We note the good agreement, obvious from Fig.~2(a), between the
full numerical solution and the result of our simple adiabatic
model. More importantly, however, we can use the analytical model
to reveal physical effects hidden in the numerically obtained $I-V$ curves. The two factors in
Eq.~(\ref{I(t)}) describe two different effects that result in a
reduction of the current (and in a negative differential
conductance) in our spintromechanical transistor. The first factor
in Eq.~(\ref{I(t)}), when averaged over time, describes the
current suppression that originates from the increased contact
resistance of an effectively non-symmetric tunnel junction. It has
nothing to do with spin effects in our magnetic device. The second
term appears due to spin-flips in the  external magnetic field. We
see that the appearance of a voltage-dependent gap in the
Rabi-oscillations between spin-up and spin-down states could
strongly suppress the averaged current. For the parameters used in
our calculations the two factors contribute almost equally to the
current drop. This means that the spin-polaronic effects (see also
Ref.~\onlinecite{pulkin}) are significant in the electric
shuttling regime in magnetic spintromechanical transistors. Note
that the time-averaged current $\bar{I}(V)$ depends on frequency
$\omega$ through the frequency dependence of the shuttling
amplitude, which can be evaluated numerically, see Fig.~2(b). We
have compared $I-V$ curves calculated by using the adiabatic model
with full numerical results for $\hbar\omega= 0.5\,\Gamma$ and
$1.5\,\Gamma$ and found agreement within an accuracy of a few per
cent (the comparison for $\hbar\omega=0.5\,\Gamma$ as well as the
curves for $\hbar\omega=1.5\,\Gamma $ are not presented in
Fig.~2(a) in order not to overload the figure with plots).

The last question we want to discuss here is: How do electron-electron
correlations influence the $I-V$ characteristics
in the transport regime where the Coulomb blockade is not pronounced?
Figure~4 shows numerical results for a Coulomb energy of
$U=4.5\,\Gamma$. All other model parameters are essentially the
same as those used for the case of noninteracting electrons (note
the different normalization current,
$e\Gamma/(2\hbar)$, as compared to Fig.~2(a)). We see that
all spin effects discussed earlier for noninteracting electrons
survive when the Coulomb interaction is turned on. There is no
qualitative change in the current-voltage dependencies. The only
new effect is the appearance of a small negative differential
conductance for the case of an immobile quantum dot (see
Ref.~\onlinecite{gorelik}).

\begin{figure}
\includegraphics[scale=0.7]{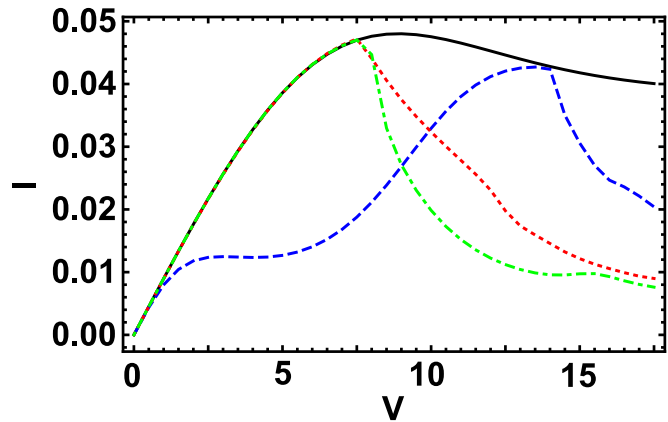}
 \caption{The $I-V$ curves of spintromechanical transistor for a Coulomb energy $U=4.5\,\Gamma$. Other parameters used in numerical
 calculations are: the level energy (counted from the Fermi energy) $\varepsilon_0/\Gamma=1$, magnetic field $\Omega_H/\Gamma=0.1$,
 temperature $T/\Gamma=1.5$, exchange coupling $J/\Gamma \simeq 0.8$,
 the electromechanical coupling constant $\kappa=0.06$ and the distance between the leads is set to be $d=20\lambda$.
 Note that the current is normalized to $(e\Gamma)/(2\hbar)$.
 The dashed curve (blue on-line) corresponds to $\hbar\omega/\Gamma\simeq 0.16$, the short-dashed curve (red on-line) - $\hbar\omega/\Gamma= 0.5$, the dashed-dotted curve (green on-line) - $\hbar\omega/\Gamma= 1.5$ and the solid curve (black
 on-line) represents current-voltage dependence for the spintronic device with a non-movable quantum dot.}
 \end{figure}

In summary, we have shown that spin-polaronic effects give rise to special features in the current-voltage characteristics
of a spintromechanical magnetic transistor even in the case when exchange forces do not lead to magnetic shuttling.
Both the low-voltage (vibronic) and the high-voltage (electric shuttling) phases demonstrate unusual behavior, which
is characterized by a differential conductance with a non-monotonic voltage dependence (vibronic phase) and a negative
differential conductance (electric shuttling phase). We predict that the appearance of
a negative differential conductance can be used as a signature of shuttling in
spintromechanical magnetic devices.


\textbf{\emph{Acknowledgement}}. This work was supported by the
Institute for Basic Science in Korea (IBS-R024-D1); the National
Academy of Sciences of Ukraine (grant No. 4/19-N and Scientific
Program 1.4.10.26.4); the Croatian Science Foundation, project
IP-2016-06-2289, and by the QuantiXLie Centre of Excellence, a
project cofinanced by the Croatian Government and the European
Union through the European Regional Development Fund - the
Competitiveness and Cohesion Operational Programme (Grant
KK.01.1.1.01.0004). The authors acknowledge the hospitality of PCS
IBS in Daejeon (Korea). OAI thanks V.V.~Slavin and Y.V.~Savin for
the help in organization of computer calculations.

\end{document}